\documentclass[prd,preprint,superscriptaddress,showpacs,amsmath,amssymb]{revtex4}
\usepackage{graphicx}
\usepackage{dcolumn}
\usepackage{bm}

\begin{document}
\title{Hawking radiation as tunneling from Gravity's rainbow}
\author{Cheng-Zhou Liu}
  \email{jzhchan@yahoo.com}
  \affiliation{Department of Physics, Beijing Normal University, Beijing 100875, China}
  \affiliation{Department of Physics and Electronic Science, Binzhou College, Binzhou 256600, China}

\author{Jian-Yang Zhu }
  \email{zhujy@bnu.edu.cn}
  \affiliation{Department of Physics, Beijing Normal University, Beijing 100875, China}

\begin{abstract}
Planck scale corrections arising from deformed special relativity on Hawking
radiation in Parikh and Wilczk's tunneling framework are studied. We
calculate the emission rate of massless particles tunneling though the
corrected horizon of modified black holes from gravity's rainbow. In the
tunneling process, when a particle get across the quantum horizon, the
metric fluctuation not only due to the energy conservation but also quantum
effects of the space-time are taken into account. Our results show that, the
emission rate is related to the changes of the black hole's quantum
corrected entropy and consistent with an underlying unitary theory. In the
modified black hole, by using black hole thermodynamics, a series of quantum
correction terms including a logarithmic term to the Bekenstein-Hawking
entropy are obtained. Correspondingly, the Planck scale corrected emission
spectrum is obtained and it deviates from the thermal spectrum.
\end{abstract}

\maketitle

\section{Introduction}

Based on Hawking's great discovery \cite{1,2} that black holes have
emission of thermal radiation, the General Relativity, Quantum
Mechanics and Thermodynamics have a link with profound physical
significance and the black hole thermodynamics get a solid
fundament. However, following this, the information loss paradox
emerges \cite{3,4}. It is that, since a thermal spectrum can not
bring out any information other than only one parameter of
temperature, when a black hole evaporate away completely, all
information about the matter making up of the black hole will be
lost. Accordingly, Hawking argued that the formation and evaporation
of a black hole are not commanded by the quantum mechanics \cite{3}.
That is to say, the pure states of matter forming the black hole
evolve into the mixed states of thermal radiation, thus, the
underlying unitary theory of the quantum mechanics is violated.
Besides, in the light of Hawking's description, the black hole
evaporation is a quantum tunneling effect \cite{a}. That is to say,
due to vacuum fluctuations near the horizon, when a pair of
particles are spontaneously created just inside the horizon, the
positive energy particle can tunnel out the horizon to the infinity.
At the same time, the negative energy particle remains behind the
horizon and effectively lowers the mass of the black hole because
the negative energy orbit can not exist outside the horizon.
However, the actual derivation of Hawking radiation did not proceed
in this way, most of which based upon quantum field theory on a
fixed background space-time without considering the gravity
back-reaction of
the emitted particles and the quantum fluctuation of the space-time \cite{5,6}%
.

Recently Hawking has put forward \cite{4} his new viewpoint on the
information loss paradox that the information hidden in a black hole
could come out if Hawking radiation was not exactly thermal but had
some corrections. And that, Parikh and Wilczek have presented a
semi-classical method of calculating the emission rate by
implementing the Hawking radiation as a tunneling process from the
horizon, in which the non-thermal spectrum i.e. back-reaction
corrected radiation spectrum is obtained \cite {7,8,9}. This simply
and availably method presents the quantum tunneling description on
Hawking radiation and support information conservation in the
radiation process of black holes. In the tunneling approach, the
particles get across classically forbidden trajectories, staring
just behind the horizon onward to infinity, and the potential
barrier is created just by the particle's self-gravitation. The
crucial point of the program is that the energy conservation has
been taken into account and the background is allowed to fluctuate
because of the particle's back-reaction. Following this method, some
recent research has been dedicated to extend this tunneling study to
many cases including static or stationary black holes \cite
{10,11,12,13,14}, cosmological horizons \cite{15,16} and different
kinds of tunneling particles such as massive and charged particles
\cite{11,12}. The same results, that is, Hawking radiation is not
purely thermal spectrum, unitary theory is satisfied and information
is conserved, are obtained. However, the quantum effects of
space-time have not been considered in Parikh and Wilczk's original
work \cite{7,8,9}and much less attention in the literature was paid
to the particle's tunneling from quantum horizon \cite {17,18}.

It is remarkable that, Hawking radiation opens an important window
to quantum gravity. Meanwhile, for comprehensively understand
Hawking radiation and some closely related problems, such as the
information loss puzzle and the origin of black hole entropy, the
Planck-scale physics is necessary. In fact, in the study of black
hole thermodynamics, quantum gravity has acquired notable
achievements but it has still many challenges. For example, which is
here of interest, the derivations in string theory support the idea
that Hawking radiation can be described within a manifestly unitary
theory, but it remains a mystery how information is returned
\cite{b}. And that, the black hole emission spectrum obtained from
loop quantum gravity is still a Hawking's thermal spectrum
\cite{L1,L2}. In addition, both string theory and loop quantum
Gravity are successful in statistically explanation of black hole
entropy, but on the quantum correction to Bekenstein-Hawking (B-H)
entropy, the two leading candidate theories of quantum gravity have
a contradiction \cite{c,d,e,f,g,h}. In the two Plank-scale physics,
the coefficient of the leading-order correction has been presented
with different values. And also, even in loop quantum gravity, there
is still debate on the coefficient \cite{g,h}.

As one generally believed viewpoint, the existence of a minimally
observable length order of Planck length is a universal feature of
quantum gravity \cite {21,22}. Recently, such character has invoked
some research on the fate of Lorentz symmetry at Planck scale. The
reason is that, the character of Plank scale physics, which in
principle may contract any object to arbitrarily small size by
Lorentz boost, seemly leads to a apparent confliction with Lorentz
symmetry. At present, when keeping Planck energy as an invariant
scale, namely a universal constant for all inertial observers, to
preserve the relativity of inertial frames, a deformed formalism of
special relativity has been proposed \cite{23,24}. This deformed
special relativity (DSR) points to the possibility that the usual
energy momentum relation in special relativity may be modified in
term of the ratio of particle's energy to Planck energy. As the main
prediction of DSR, modified dispersion relations (MDR) has got both
experimental and theoretical supports \cite {24,24-1}. In the same
time, great efforts have been devoted to DSR and its implications
\cite{24,24-1,24-2,25,26,27,28,29,rain-1,30,31,31-1}. In \cite
{28,29}, based on DSR, the quantum corrections to B-H entropy have
been investigated and obtained. And, in \cite{30}, DSR has been
extended to general relativity. The main feature of DSR
incorporating with the curvature of space-time is that the geometry
of space-time depend on the energy of a particle moving in it. That
is to say, for the space-time with Planck scale correction effects
arising from DSR, there are different geometries of space-time for
particles with different energies. The modified geometries of
space-time are described by one parameter family of metric as a
function of particle energy observed by an inertial observer, namely
the rainbow metric. The Schwazschild solution of the rainbow metric
has been presented in \cite {30}. And that, it's some thermodynamics
quantities and asymptotic flatness have been investigated in
\cite{31,31-1}, respectively.

In this paper, we study the particle's tunneling from the modified
Schwazschild solution. Our aim is to incorporate Plank scale physics
with Parikh-Wikzek's semi-classical method of investigating Hawking
radiation and then to investigate some Planck scale modification
effects on black hole evaporation, entropy and information paradox.
In a general way, the black hole radiation spectrum have arbitrarily
high frequencies and their energy can go below the Planck energy
\cite{18,32,33}. Therefore, in Parikh and Wilczek' tunneling
framework, the Planck scale modification effects on the Hawking
radiation should be taken into account. In the present tunneling
investigation, while a particle tunneling though the quantum
corrected horizon, the metric fluctuation not only due to the energy
conservation but also quantum effects of the space-time is taken
into account. Here, the quantum effects of space-time are presented
by using the rainbow metric obtained from DSR. The results of the
paper show that, when the Plank scale modification of space-time is
incorporated with the particle's self-gravitation in the tunneling
program, the emission rate is related to the changes of the modified
black holes entropy and consistent with an underlying unitary
theory. Here, the black hole entropy include the quantum correction
to Bekenstein-Hawking (B-H) entropy. By using the first law of black
hole thermodynamics to the modified black holes, a series of quantum
correction terms including a logarithmic term are obtained and the
result is consistent with a rigorously statistical calculation in
loop quantum gravity \cite{h}. Correspondingly, the emission
spectrum with correction arising from energy conservation also
quantum gravity effects is obtained. The corrected spectrum is
departure from the pure thermal spectrum and it's statistical
correlations is discussed.

The paper is organized as follows. In Sec. \ref{Sec. 2}, the modified
Schwarzschild solution from the gravity's rainbow in the context of DSR is
introduced briefly and its some thermodynamics quantities is investigated.
Then in Sec.\ref{Sec. 3}, by using the Parikh-Wikzek's framework, the
emission rates of massless particles tunneling through the horizon of the
modified black holes are calculated. Then in Sec.\ref{Sec. 4}, the entropy
of the modified black hole is investigated and the quantum correction
include a logarithmic item to the B-H entropy is obtained. Accordingly, the
deviation of the emission spectrum of the modified black hole to the thermal
spectrum is obtained. The last part is the summary and conclusion.

\section{The modified black holes from gravity's rainbow}

\label{Sec. 2}

To investigate the Hawking radiation as tunneling from a quantum corrected
horizon, firstly we briefly introduce the modified Schwarzschild solution
from gravity's rainbow and analyze its some thermodynamics quantities.

The staring point and main result of DSR is MDR \cite{24}, namely
\begin{equation}
E^2f_1^2\left( E{;\lambda }\right) -p^2f_2^2\left( E{;\lambda }\right)
=m_0^2,  \label{eq1}
\end{equation}
where $f_1$and $f_2$ are two energy functions from which a specific
formulation of boost generator can be defined, in which $\lambda $ is a
parameter of order the Planck length. The equation concretely indicates
that, MDR is energy dependent. It is to say, particles with different energy
$E$ have different energy-momentum relations.

From DSR, it has been pointed out that the flat space-time has the invariant%
\cite{rain-1,30}
\begin{equation}
ds^2=-\frac{dt^2}{f_1^2}+\frac{dr^2}{f_2^2}+\frac{r^2}{f_2^2}d\Omega ^2.
\label{eq2}
\end{equation}
Thus, the DSR space-time is endowed with an energy dependent quadratic
invariant, that is, an energy dependent metric, namely rainbow metric.

By extending the Eq.(\ref{eq2}) to incorporate curvature, in \cite{30}, the
modified Schwarzschild solution has been demonstrated in terms of energy
independent coordinates and the energy independent mass parameter $M$ as
\begin{equation}
dS^2=-\frac{\left( {1-}\frac{2GM}r\right) }{f_1^2}dt_s^2+\frac 1{f_2^2\left(
{1-}\frac{2GM}r\right) }dr^2+\frac{r^2}{f_2^2}d\Omega ^2.  \label{eq3}
\end{equation}
Obviously, the metric is also depend on the energy of particle moving in it.
That is, if a given observer probes the space-time using the quanta with
different energies, he will conclude that space-time geometries have
different effective description. Here, the particle's energy $E$ denotes
it's total energy measured at infinity from the black hole. By this, the
present space-time is endowed with a Plank-scale modification and has some
quantum effects.

In addition, from \ref{eq3}, we can see the modified Schwarzschild solution
is asymptotically DSR. And that, it has been pointed that the asymptotically
DSR space-times has equality with the usual asymptotically flat space-times
\cite{31-1}. Then, using the Komar integrals, we define the total
Arnowitt-Deser-Misner (ADM) mass $M_{ADM}$ of the quantum effected
space-time as
\begin{equation}
M_{ADM}=-\frac 1{8\pi G}{\int_s{\varepsilon _{abcd}\nabla ^c\xi ^d}}=\frac M{%
f_1f_2}.  \label{eq4}
\end{equation}
We find that, for the modified Schwarzschild black holes, the ADM mass is
not equal to the mass parameter $M$. It shows that the quantum corrected
space-time have topological defects. And that, the total energy of the
space-time is depend on the energy of the probe particle. This is because
the quantum correction effects of the black hole arise from DSR, which has
energy dependence.

Also, From the metric (\ref{eq3}), we can see its horizon $r_{+}=2GM$ is
universal for all observers and at the usual place as the usual
Schwarzschild black hole. However, the horizon area
\begin{equation}
A=\frac{16\pi G^2M^2}{f_2^2}  \label{eq6}
\end{equation}
is different from the usual value and depend on the particle's energy. This
should has some modification effects on the black hole thermodynamics.

Besides, the surface gravity on the horizon is defined by
\begin{equation}
\kappa =-\frac 12{\lim_{r-r_{+}}}\sqrt{\frac{-g^{rr}}{g^{tt}}}\frac{\left( {%
g^{tt}}\right) ^{\prime }}{g^{tt}}.  \label{eq7}
\end{equation}
Thus, from (\ref{eq3}), we can obtain the surface gravity of the modified
black hole as
\begin{equation}
\kappa =\frac{f_2}{f_1}\frac 1{4GM}.  \label{eq8}
\end{equation}
Obviously, it also depends on the particle's energy.

Therefore, the temperature of the modified black hole is obtained as
\begin{equation}
T=\frac \kappa {2\pi }=\frac{f_2}{f_1}\frac 1{8\pi GM}.  \label{eq9}
\end{equation}
It show that the temperature of the modified Schwarzschild solution is
dependent on the energy of probe particle. That is, using the quanta with
different energy, an observer at infinity will probe different effective
temperature for the quantum corrected black hole.

Obviously, in the modified black holes, the energy dependence of
thermodynamics quantities arises from DSR and is the exhibition of quantum
effects of the space-time. And that, analysis on the characters is necessary
for the applications of energy conservation and black hole thermodynamical
law in the quantum corrected space-time.

\section{Tunneling probability in modified black holes}

\label{Sec. 3} DSR and gravity's rainbow are low energy effect of
quantum gravity. It is that, the modified black holes (\ref{eq3}) is
a coarse grained model of space-times at semi-classical level. Here,
we assume that Parikh and Wilczk's quantum tunneling program of
investigating Hawking radiation is still holds for the large
modified black holes. Therefore, in this section, following Parikh
and Wilczk's tunneling framework, we calculate the tunneling
probability of massless particles in the quantum modified black
hole. The novel point of the present tunneling investigation is that
the quantum effect of geometry are considered in the tunneling
process.

In Parikh and Wilczk's tunneling scheme, the particle behind the horizon can
tunnel out along a classically forbidden trajectory and the tunneling
probability is given by means of WKB approximation. That is, the emission
rate can be expressed as the imaginary part of the action for the trajectory
\cite{7,8,9}
\begin{equation}
\Gamma \sim \exp (-2%
\mathop{\rm Im}%
I).  \label{eq10}
\end{equation}

For calculating the action $I$ in the modified black holes showing as (\ref
{eq3}), the coordinate singularity at the horizon must be removed. Here, we
introduce a new time coordinate $t$ and follow a Painleve type coordinate
transformation. Letting
\begin{equation}
dt_s=dt-F\left( r\right) dr,  \label{eq11}
\end{equation}
and
\begin{equation}
\frac 1{f_2^2\left( {1-}\frac{2GM}r\right) }-\frac{\left( {1-}\frac{2GM}r%
\right) }{f_1^2}F^2\left( r\right) =1,  \label{eq12}
\end{equation}
then we have
\begin{equation}
ds^2=-\frac{\left( {1-}\frac{2GM}r\right) }{f_1^2}dt^2+\frac 2{f_1f_2}\sqrt{%
1-f_2^2\left( {1-}\frac{2GM}r\right) }dtdr+dr^2+\frac{r^2}{f_2^2}\left( {%
d\theta ^2+\sin ^2\theta d\varphi ^2}\right) .  \label{eq13}
\end{equation}
It is easy to find that the Painleve-like metric of the modified black holes
has some advantages for us to implement the calculation on the emission rate
of particle tunneling through a quantum corrected horizon. Firstly, none of
the components of either the metric or the inverse metric diverges at the
horizon. Secondly, the coordinate system has Killing vector ${\partial }/{%
\partial t}$. In addition, as expected, the metric has Planck scale
correction effects showing as the energy dependence. It denotes that, even
if the black hole has a fixed mass parameter $M$, the emitted particles with
different energy will be affected by different metric.

It is assumed that a massless particle with energy ${E}=\frac 1{f_1f_2}{%
\omega }$ measured at infinity tunnels out the horizon of the modified black
hole. For the massless particle, its motion equation can be given by the
radial null geodesics on the geometry (\ref{eq13}). Let $ds^2=0$, in the
presence of Planck scale effects, we have the radial null geodesic as
\begin{equation}
\dot{r}=\frac{dr}{dt}=\frac 1{f_1f_2}\left[ {\pm 1-}\sqrt{1-f_2^2\left( 1-%
\frac{2GM}r\right) }\right] ,  \label{eq14}
\end{equation}
where ''+'' corresponding outgoing particles, ''-''corresponding ingoing
particles.

However, if we enforce the energy conservation of the space-time,
when the particle tunnels out the horizon, the mass of the black
hole should vary. That is, the back-reaction of emitted particles
should affect the background geometry. In spherical symmetry
space-time, the back-reaction effects of emitted shell have been
investigated in detail \cite{35}. Here, we treat the particle as a
s-wave i.e. a shell. Thus, when the particle radiate outside the
horizon of the modified black hole, since the particle's
self-gravitation, the mass parameter $M$ in the metric (\ref{eq13})
should be replaced with $M-{\omega }$\cite{35,7,8,9}. This is
consistent with Birkhoff's theorem. The theorem tell us that, in the
spherical symmetry space-time, the only effect on the geometry due
to the s-wave is to provide a junction condition for matching the
total mass inside and outside the shell. Then we get the geometry
between the horizon and the spherical shell as
\begin{equation}
ds^2=-\frac{\left( {1-}\frac{2G\left( M-\omega \right) }r\right) }{f_1^2}%
dt^2+\frac 2{f_1f_2}\sqrt{1-f_2^2\left( {1-}\frac{2G\left( M-\omega \right) }%
r\right) }dtdr+dr^2+\frac{r^2}{f_2^2}\left( {d\theta ^2+\sin ^2\theta
d\varphi ^2}\right) .  \label{eq15}
\end{equation}
And that, we can see the locations of the horizon before and after the
particle's emission are $r_i=r_{+}\left( M\right) =2GM$ and $r_f=r_{+}\left(
M{-\omega }\right) =2G\left( M-\omega \right) $, respectively. Thus, by the
shrinking of the black hole, the tunneling barrier is created by the emitted
particle itself due to the energy conservation of the space-time.
Furthermore, the set of the barrier is not affected by the Planck scale
modification effects of the black hole-emitted particle system.

In addition, it has been proved that the motion of the emitted
particle is affected by the geometry between the event horizon and
the spherical shell \cite{35}. Thus, considering the background
metric's dynamical effects due to energy conservation and the
quantum effects of the space-time, the tunneling particle's radial
motion equation should be modified as
\begin{equation}
\dot{r}=\frac{dr}{dt}=\frac 1{f_1f_2}\left( {1-}\sqrt{1-f_2^2\left( 1-\frac{%
2G\left( M-\omega \right) }r\right) }\right) .  \label{eq16}
\end{equation}

Also, for the tunneling process, a canonical Hamiltonian treatment gives a
simple result for the total action of the black hole-particle system \cite
{35},
\begin{equation}
I=\int dt\left( p_t+\frac{dr}{dt}p_r\right) ,  \label{eq17}
\end{equation}
where $p_t$ and $p_r$ are the conjugate momentum corresponding to Painleve's
coordinates $t$ and $r$, respectively. Here, only the second term in (\ref
{eq17}) contributes to the imaginary part of the action. Therefore, we can
obtain the tunneling probability for an outgoing massless particle by
computing the imaginary part of the action, which is
\begin{equation}
\mathop{\rm Im}%
I=%
\mathop{\rm Im}%
\int\limits_{t_i}^{t_f}dt\frac{dr}{dt}p_r=%
\mathop{\rm Im}%
\int\limits_{r_i}^{r_f}p_rdr=%
\mathop{\rm Im}%
\int\limits_{r_i}^{r_f}\int\limits_0^{p_r}dp_r^{\prime }dr,  \label{eq18}
\end{equation}
where $t_i$ and $t_f$ are the Painleve coordinate times corresponding $r_i$
and $r_f$, respectively.

To proceed with an explicit computation, we now apply the Hamilton's
equation
\begin{equation}
\stackrel{\cdot }{r}=\frac{dH}{dp_r}=\frac{dM_{ADM}^{\prime }}{dp_r},
\label{eq19}
\end{equation}
there $M_{ADM}^{\prime }=\frac 1{f_1f_2}M^{\prime }$is the ADM mass of the
modified black hole after emitting a particle with energy $E^{\prime }=\frac %
1{f_1f_2}{\omega }^{\prime }$. Substituting Eq.(\ref{eq19}) into Eq.(\ref
{eq18}), and switching the order of integral, we have
\begin{eqnarray}
\mathop{\rm Im}%
I &=&%
\mathop{\rm Im}%
\int\limits_{r_i}^{r_f}\int\limits_0^{p_r}\frac{dM_{ADM}^{\prime }}{%
\stackrel{\cdot }{r}}dr{{=%
\mathop{\rm Im}%
}}\int\limits_M^{M-\omega }dM_{ADM}^{\prime }\int\limits_{r_i}^{r_f}\frac{dr%
}{\stackrel{\cdot }{r}}  \nonumber \\
&=&%
\mathop{\rm Im}%
\int\limits_M^{M-\omega }\int\limits_{r_i}^{r_f}\frac{f_1}{f_2}\frac{r\left(
1+\sqrt{1-f_2^2\left( 1-\frac{r_{+}^{\prime }}r\right) }\right) }{%
r-r_{+}^{\prime }}drd\frac{M^{\prime }}{f_1f_2},  \label{20}
\end{eqnarray}
where $r_{+}^{\prime }=2GM^{\prime }$ is the horizon location after emitting
the particle. Considering the particle tunneling through the horizon, we can
see that $r_{+}^{\prime }$ is a single pole in Eq.(\ref{20}). Then the
integral can be evaluated by deforming the contour around the pole. In this
way, we finished the integral over $r$ and get
\begin{equation}
\mathop{\rm Im}%
I=-4\pi G\int\limits_M^{M-\omega }\frac{f_1}{f_2}M^{\prime }d\frac{M^{\prime
}}{f_1f_2}.  \label{eq21}
\end{equation}
Now, for the modified black hole in the tunneling process, we apply the
first law of black hole thermodynamics
\begin{equation}
dM_{ADM}^{\prime }=T^{\prime }d{S}^{\prime }.  \label{eq22}
\end{equation}
In fact, many previous works \cite{12,13,14} in Parikh and Wilczk's
tunneling framework have confirmed that the tunneling process is consistent
with the first law of black hole thermodynamics. Then, inserting the
temperature expression (\ref{eq9}) into (\ref{eq22}), we have
\begin{equation}
4\pi G\frac{f_1}{f_2}M^{\prime }d\frac{M^{\prime }}{f_1f_2}=\frac 12d{S}%
^{\prime },  \label{eq23}
\end{equation}
and
\begin{equation}
\mathop{\rm Im}%
I=-\frac 12\int\limits_S^{S+\triangle S}dS^{\prime }=-\frac 12\Delta S,
\label{eq24}
\end{equation}
where $\Delta S=S\left( M-{\omega }\right) -S\left( M\right) $ is the
difference of the black hole entropies of the modified black hole before and
after the emission.

Thus the tunneling probability of the massless particle from the quantum
corrected horizon is obtained, namely
\begin{equation}
\Gamma =\exp \left( {-2%
\mathop{\rm Im}%
I}\right) =\exp \left( {\Delta S}\right) .  \label{eq25}
\end{equation}

We find that the tunneling rate is related to the change of the modified
black hole entropy and is consist with an underlying unitary theory. It is
the same result obtained from the usual Schwarzschild black hole \cite{7,8,9}%
. However, in the quantum corrected space-time, the present black hole
entropy should have quantum correction to B-H entropy. This is a radical
difference with Parikh and Wilczk's original results, in which, black hole
entropy is obtained and applied as B-H entropy. Accordingly, the emission
spectrum of the modified black hole should has corresponding Planck scale
corrections to the usual spectrum from the usual black hole. In the next
section, by calculating the (\ref{eq23}), we obtain the quantum corrected
entropy and the corrected emission spectrum of the modified black hole is
given and discussed.

\section{Entropy and radiation spectrum of the modified black holes}

\label{Sec. 4} In the present tunneling investigation, to obtain the
entropy of the modified black hole by calculating the (\ref{eq23})
and based on the entropy to analyze the radiation spectrum, we need
the explicit DSR i.e. specific correction functions $f_1$ and $f_2$.
Some research has been devoted to the
investigation on the explicit MDR models and different functions $f_1$ and $%
f_2$ have been proposed \cite{24,24-2}. However, as so far, the
standard form of $f_1$ and $f_2$ has not been given and the further
investigation are necessary. In the low energy realm i.e.
$E/E_p{<<1}$, where $E_p\equiv 1/ \sqrt{8\pi G}$ is the Planck
energy, the correspondence principle requires that $f_1$ and $f_2$
approach to unit. Here, for convenience, we take
\begin{equation}
f_1=f=e^{-\frac 12E^2/E_p^2},f_2=1.  \label{eq33}
\end{equation}

Then, based on the specific MDR, from the Eq.(\ref{eq6}) and Eq.(\ref{eq9}),
the horizon area and the temperature of the modified black holes are,
respectively,
\begin{equation}
A=16\pi G^2M^2,  \label{eq34}
\end{equation}
\begin{equation}
T^2=\frac 1{f^2}\frac 1{\left( {8\pi GM}\right) ^2}.  \label{eq35}
\end{equation}
And that, from Eq.(\ref{eq23}), we have the differential equation of the
black hole entropy as
\begin{equation}
dS=8\pi GMfd\left( \frac Mf\right) =4\pi GdM^2-8\pi GM^2\frac{df}f.
\label{eq36}
\end{equation}
We find that, for the modified black hole, the entropy equation is
depend on the energy of particle. It is to say, the effective black
hole entropy has dependence on the energy of probe particle. Now,
for large modified black holes, we use the characteristic
temperature by identifying the energy of particles emitted from the
black holes with the hole's temperature \cite {36,37,29,31}, namely
\begin{equation}
E=T.  \label{eq37}
\end{equation}
This can be understood as a statistical treatment for explicitly obtaining
the black hole entropy. It is that, supposing all the emitted particles form
an ensemble outside the black hole, then the average energy of the particles
is equal to the temperature of the black hole. In other words, we us the
particle with energy $T$ to probe the entropy and ascertain it as the
intrinsic entropy i.e. the black hole entropy. Then we have
\begin{equation}
f=e^{-\frac 12T^2/E_p^2},\frac{df}f=-\frac 1{2E_p^2}dT^2.  \label{eq38}
\end{equation}
For the large modified black hole with $M>>1/\sqrt{8\pi G}$, we have $%
T^2/E_p^2\sim E_p^2/M^2<<1$, and
\begin{equation}
T^2=\frac 1{f^2}\frac 1{\left( {8\pi GM}\right) ^2}=\left( 1+\frac{T^2}{E_p^2%
}+\cdots \right) \frac 1{\left( {8\pi G}\right) ^2M^2}\simeq \frac 1{\left(
8\pi G\right) ^2M^2}+\frac 1{8\pi GM^2}T^2.  \label{eq40}
\end{equation}
Substituting Eq.(\ref{eq34}) into Eq.(\ref{eq40}) and solving it, we have
\begin{equation}
T^2=\frac 1{4\pi A}\frac 1{1-\frac{2G}A}=\frac 1{4\pi A}\left( 1+\frac{2G}A%
+\left( \frac{2G}A\right) ^2+\cdots \right) =\frac 1{4\pi A}+\frac 1{4\pi A}%
\sum\limits_{n=1}^\infty \left( \frac{2G}A\right) ^n.  \label{eqyy}
\end{equation}
Then, substituting Eq.(\ref{eq38}) and Eq.(\ref{eqyy}) into the Eq.(\ref
{eq36}), we obtain
\begin{equation}
dS=d\left( {\frac A{4G}}\right) -\frac 12d\ln \left( {\frac A{4G}}\right)
+\sum\limits_{n=1}^\infty \frac{n+1}n\frac 1{2^{n+1}}d\left( \frac{4G}A%
\right) ^n.  \label{eq41}
\end{equation}
Next, integrating the Eq(\ref{eq41}), up to a constant term, the entropy
expression of the modified black holes can be obtained as
\begin{equation}
S=\frac A{4G}+c_0\ln \left( \frac A{4G}\right) +\sum\limits_{n=1}^\infty
c_n\left( \frac{4G}A\right) ^n+const,  \label{eq43-1}
\end{equation}
where $c_0=-1/2$, $c_n=\left( n+1\right) /n2^{n+1}$.

It is worth to point out that, the present leading order correction to the
B-H entropy goes as the logarithm of the black hole area. It is consistent
with many other research's results (for a review of the correspondence see%
\cite{m-1}). In particle, the factor $c_0=-1/2$ is the same as the
loop quantum theory prediction by the direct counting the
micro-states of black holes \cite{h}. Of course, this happen on the
condition that a special form of MDR shown as Eq.(\ref{eq33}) has
been proposed.

Now, substituting Eq.(\ref{eq43-1}) into the Eq.(\ref{eq25}) and thinking of
the Eq(\ref{eq34}), we can obtain the radiation spectrum of the modified
black holes, namely,
\begin{eqnarray}
\Gamma &\sim &\exp \left( \triangle S\right) =\exp \left( S(M-\omega
)-S\left( M\right) \right)  \nonumber \\
&=&\left[ \prod\limits_{n=1}^\infty \exp \left( \frac{c_n}{\left( 4\pi
G\right) ^n}\frac{1-\left( 1-\frac \omega M\right) ^{2n}}{M^{2n}\left( 1-%
\frac \omega M\right) ^{2n}}\right) \right] \left( {1-}\frac \omega M\right)
^{-1}\exp \left( -8\pi GM\omega \left( 1-\frac \omega {2M}\right) \right) .
\label{43-2}
\end{eqnarray}

Compared with the usual self-gravitation correction radiation spectrum from
the usual black holes derivation found in \cite{7,8,9}
\begin{equation}
\Gamma \sim \exp \left( -8\pi GM\omega \left( 1-\frac \omega {2M}\right)
\right) ,  \label{eqy1}
\end{equation}
we find the present radiation spectrum has a series of quantum modification
factors and it further derived from the pure thermal spectrum. But, If we do
not consider the corrections from quantum effects of gravity, i.e.
neglecting the logarithmic correction term and the inverse area items in the
black hole entropy, the factor of the final exponential in Eq.(\ref{43-2})
equals to unit and the radiation spectrum has the same type of non-thermal
form shown as (\ref{eqy1}). And that, if we further overlook the effect of
emitted particle's back-reaction with neglecting $\omega /M$ in the
expression (\ref{43-2}), the tunneling rate of the large modified black
holes takes the form of Boltzmann factor $e^{-\beta \omega }$ with $\beta
\equiv 1/T=8\pi GM$ and the Hawking's thermal formula is obtained.

Next, for analyzing how to get information from the radiation
spectrum, we discuss the statistical correlations for the
probabilities of different emission models in the present emission
spectrum \cite{9,17,18}. The statistical correlation between two
emission probabilities of two emitted quanta is measured by the
function
\begin{equation}
\chi \left( {\omega _1+\omega _2;\omega _1,\omega _2}\right) =\ln (\Gamma
(\omega _1+\omega _2))-\ln (\Gamma (\omega _1)\Gamma (\omega _2)).
\label{eq44}
\end{equation}
Where we assumes that the two quanta with energies $\omega _1$ and $\omega
_2 $ are emitted out successively and $\Gamma \left( \omega _1+\omega
_2\right) $ presents the tunneling probability of particle with energy $%
\omega =\omega _1+\omega _2$. For the Hawking's thermal emission spectrum,
the correlation function is zero showing that the probabilities of different
models are independent. And that, for the back-reaction corrected
non-thermal spectrum in the usual black hole, the function also vanishes\cite
{9}. It is that, in Parikh-Wikzek's tunneling program, the back-reaction
effects alone do not provide a statistical correlations between different
quanta in Hawking radiation.

From Eq.(\ref{43-2}), we have the emission rates of the two particles of $%
\omega _1$and $\omega _2$ as, respectively,
\begin{equation}
\ln [\Gamma \left( {\omega _1}\right) ]=-8\pi GM\omega _1\left( {1-}\frac{%
\omega _1}{2M}\right) -\ln \left( 1-\frac{\omega _1}M\right)
+\sum\limits_{n=1}^\infty \frac{c_n}{\left( 4\pi G\right) ^n}\frac{1-\left(
1-\frac \omega M\right) ^{2n}}{M^{2n}\left( 1-\frac{\omega _1}M\right) ^{2n}}%
,  \label{eq45}
\end{equation}
\begin{eqnarray}
\ln [\Gamma (\omega _2)] &=&-8\pi G(M-\omega _1)\omega _2\left( {1-}\frac{%
\omega _2}{2\left( M-\omega _1\right) }\right)  \nonumber \\
&&-\ln \left( 1-\frac{\omega _2}{M-\omega _1}\right)
+\sum\limits_{n=1}^\infty \frac{c_n}{\left( 4\pi G\right) ^n}\frac{1-\left(
1-\frac{\omega _2}{M-\omega _1}\right) ^{2n}}{\left( M-\omega _1\right)
^{2n}\left( 1-\frac{\omega _2}{M-\omega _1}\right) ^{2n}}.  \label{eq46}
\end{eqnarray}
Alternatively, a single emission of the particle with energy $\omega $ is
\begin{eqnarray}
\ln [\Gamma (\omega _1+\omega _2)] &=&-8\pi GM(\omega _1+\omega _2)\left( {1-%
}\frac{\omega _1+\omega _2}{2M}\right)  \nonumber \\
&&-\ln \left( 1-\frac{\omega _1+\omega _2}M\right) +\sum\limits_{n=1}^\infty
\frac{c_n}{\left( 4\pi G\right) ^n}\frac{1-\left( 1-\frac{\omega _1+\omega _2%
}M\right) ^{2n}}{M^{2n}\left( 1-\frac{\omega _1+\omega _2}M\right) ^{2n}}.
\label{eq47}
\end{eqnarray}
Thus, we obtain the correlation function of the two tunneling models as
\begin{equation}
\chi (\omega _1+\omega _2;\omega _1,\omega _2)=0.  \label{eq48}
\end{equation}

We find that, for the quantum corrected black hole, the correlation function
between two sequential emission rate still vanishes as in the usual black
hole. That is to say, the Planck scale corrected non-thermal spectrum
obtained in the modified black hole, up to high order correction factors,
has not provided the clear mechanism for recovery of information from the
tunneling process. It implies that, how to straightway obtain information
from the emission rate of black hole is still an open problem.

\section{Summary and Discussions}

\label{Sec. 5}

In the present work, incorporating Planck scale physics with the
Parikh-Wikzek's tunneling program \cite{7,8,9}, Hawking radiation as
tunneling of particles in the modified black hole from the rainbow
metric is investigated. While the particles tunnel across the
horizon, by applying DSR and it's quantum space-time effects, the
quantum gravity effects of the black hole-particles system are taken
into account. Thus, both the geometry of space-time and the energy
momentum relation of particle depend on the particle's energy.
Therefore, in the tunneling process from the quantum corrected black
hole, the background metric is dynamical, due to not only energy
conservation but also the quantum effects of geometry. We find that,
in the context of DSR and gravity's rainbow, the tunneling
probabilities of massless particle from the quantum corrected
horizon are related to the changes of the modified black holes
entropy, and the derived emission spectrum depart from the pure
thermal spectrum, but it is consistent with an underlying unitary
theory.

For the modified black hole, by analyzing it's some thermodynamics
quantities and using the first law of black hole thermodynamics, the
entropy with a series of quantum correction terms is obtained. Here,
the leading order correction item is the logarithm of the black hole
area and the expression of black hole entropy is consistent with a
statistical calculation in loop quantum gravity \cite{h}.
Accordingly, the Planck scale corrected emission spectrum in the
modified black hole is obtained and it deviates from the thermal
spectrum. However, though the emission spectrum present a series of
quantum correction factors, it has not provide a desired correlation
between different emission models. Then, in the quantum corrected
black hole, how to encode information and obtain it by the tunneling
process is not as clear as in the usual black hole. Meanwhile, from
the entropy calculation in the modified black hole, a specific MDR
of Eq(\ref{eq33}) is selected and the obtained result of entropy
formula Eq(\ref {eq43-1}) support the choice. Black hole entropy is
an important landmark to Planck scale physics. As a low-energy
quantum gravity effect, different models of MDR with some underlying
meaning in the quantum gravity should be tested with black hole
entropy.

The research here not only provides further evidence to support the
Parikh-Wikzek's tunneling program, which gives an explicit calculation to
investigate Hawking radiation and related problem, but also gives an
extension for the tunneling analysis from classical black hole to a quantum
corrected black hole. And that, the work can be extended to other modified
black holes from gravity's rainbow. On this issue, further work is in
progress.

\acknowledgments The work was supported by the National Natural
Science of China (No.10375008) and the National Basic Research
Program of China (2003CB716302).

\end{document}